\documentclass[manuscript]{acmart}
\AtBeginDocument{%
  \providecommand\BibTeX{{%
    \normalfont B\kern-0.5em{\scshape i\kern-0.25em b}\kern-0.8em\TeX}}}

\setcopyright{acmcopyright}
\copyrightyear{2023}
\acmYear{2023}
\acmDOI{XXXXXXX.XXXXXXX}

\acmConference[FashionXRecSys ’23]{ACM Conference on Recommender Systems}{September 18, 2023}{Singapore}
\acmBooktitle{Singapore '23: ACM Conference on Recommender Systems (FashionXRecsys Workshop),
 September 18, 2023, Singapore} 
\acmPrice{15.00}
\acmISBN{978-1-4503-XXXX-X/18/06}

\usepackage{multirow}
\usepackage{paralist}
\usepackage{subfig}
\usepackage{adjustbox}

\usepackage{xcolor}
\usepackage{listings}

\lstdefinestyle{yaml}{
     basicstyle=\color{blue}\footnotesize,
     rulecolor=\color{black},
     string=[s]{'}{'},
     stringstyle=\color{blue},
     comment=[l]{:},
     commentstyle=\color{black},
     morecomment=[l]{-}
 }

\begin{document}

\title{Tailor: Size Recommendations for High-End Fashion Marketplaces}

\author{Alexandre Candeias}
\affiliation{
    \institution{FARFETCH}
    \city{Lisbon}
    \country{Portugal}
}
\email{alexandre.candeias@farfetch.com}

\author{Ivo Silva}
\affiliation{
    \institution{FARFETCH}
    \city{Porto}
    \country{Portugal}
}
\email{ivomiguel.silva@farfetch.com}

\author{Vitor Sousa}
\affiliation{
  \institution{FARFETCH}
  \city{Porto}
  \country{Portugal}
}
\email{vitor.costasousa@farfetch.com}

\author{José Marcelino}
\affiliation{
  \institution{FARFETCH}
  \city{London}
  \country{United Kingdom}
}
\email{jose.marcelino@farfetch.com}

\renewcommand{\shortauthors}{Candeias, et al.}

\begin{abstract}
In the ever-changing and dynamic realm of high-end fashion marketplaces, providing accurate and personalized size recommendations has become a critical aspect. Meeting customer expectations in this regard is not only crucial for ensuring their satisfaction but also plays a pivotal role in driving customer retention, which is a key metric for the success of any fashion retailer. \\
We propose a novel sequence classification approach to address this problem, integrating implicit (\textit{Add2Bag}) and explicit (\textit{ReturnReason}) user signals.
Our approach comprises two distinct models: one employs LSTMs to encode the user signals, while the other leverages an Attention mechanism.
Our best model outperforms SFNet, improving accuracy by 45.7\%.
By using \textit{Add2Bag} interactions we increase the user coverage by 24.5\% when compared with only using \textit{Orders}.
Moreover, we evaluate the models' usability in real-time recommendation scenarios by conducting experiments to measure their latency performance.
\end{abstract}

\ccsdesc[500]{Computing methodologies~Machine learning approaches}

\keywords{Size Recommendation, Recommendations Systems, Size and Fit, Personalization, E-commerce}



\maketitle

\section{Introduction}

With the increase in online shopping for fashion over the last years \cite{stateecommerceshopify2023}, customers are more comfortable purchasing garments without any physical contact.
In the fashion industry, this leads to an increase in returns related to Size\&Fit reasons which bring both financial and environmental losses.

Different strategies are available to mitigate this problem from an online perspective.
Product size-related information available like fitting advice, size tables, and product measurements helps customers when deciding which size to order.
However, it lacks customer-specific personalization and size choice is not only dependent on the product but also on the customer preferences.
Questionnaires, where the customer inserts data about height, weight, and body shape, can be used to gather customer information and preferences.
But it can also be seen as intrusive by customers, and more importantly, this kind of information varies over time and would need to be updated.

Lately, the usage of virtual try-on technologies is also getting some adoption in fashion marketplaces.
Even though in clothing articles it still lacks the necessary precision and real feeling to be used more broadly.
The customer might also feel intrusive to share image data of his body.

An alternative that requires less customer adoption and can be used in all marketplaces is to use historical interactions between customers and products, for example, past purchases.
This data can be used to make specific size recommendations for each pair (customer, product). In this work, we explore how this approach can be used in a luxury fashion marketplace.
In the rest of this paper, when we refer to size recommendation we mean the problem of predicting a size for a specific (customer, product) pair.

Luxury fashion marketplaces face some particular challenges when compared to regular fashion marketplaces.
The products are premium and have a limited size availability.
This makes that a wrong size ordered from one client might represent the end of stock for that specific size of the product.
The return process also has some particular issues, due to the high value of the product, returns are more complicated which leads to client satisfaction impacts.

To add complexity, in luxury fashion, the number of products available and brands is usually higher than in normal fashion.
This translates into a bigger sparsity of interactions between the user and the products and a bigger number of available sizes.

In this work, we propose "Tailor" a solution to the size recommendation problem in a luxury fashion marketplace that uses only historical transactional data.
Our contributions are the following:

\begin{enumerate}
    \item Findings regarding size recommendations within a high-end fashion marketplace, accompanied by an evaluation of how existing baselines perform when faced with complex data sets.
    \item Introduce two models that make use of additional customer signals such as Add2Bag interactions and feedback from returned orders.
    \item Analysis of the models' latency, to validate the usability in real-time recommendations scenarios.
\end{enumerate}

\section{Related Work}
In recent times, there has been a significant surge of academic interest, particularly from the industry, in the Size\&Fit domain.

In \cite{Nestler2021,Karessli2019} product-level advising using visual and transactional data is proposed. 
These approaches can be very efficient in terms of the coverage of users since they don't require a user to have interactions with the marketplace, solving the cold start problem.
However, this type of recommendation lacks personalization to the customer-specific needs, and they are often generic such as: "This product usually fits larger than usual" or  "This product usually fits smaller than usual".

Virtual try-on targeting Size\&Fit customer experience is proposed in \cite{Lewis2022, Pecenakova2022}. 
These works focus more on the fitting and visual apparel of the garment in the customer and don't provide any specific size recommendation.

In the size recommendation area, many works have been proposed: \cite{Sembium2018,Guigoures2018,Sheikh2019,Hajjar2021,Eshel2021,Dogani2019,MohammedAbdulla2019,Lasserre2020,Singh2018,Abdulla2017}.
Bayesian approaches \cite{Guigoures2018,Sembium2018}, while simple and scalable to large datasets were surpassed by embedding-based models \cite{Sheikh2019}.
Embedding-based approaches \cite{Abdulla2017, Sheikh2019, Dogani2019} are usually focused on learning embedding representations for the customer and the product.
At inference time, these representations can be used to make predictions however, we are often limited to the customers or products that were used at training time.
To overcome the dependency on training time, many sequence-based models have been proposed \cite{MohammedAbdulla2019,Eshel2021,Hajjar2021}.
These models typically see the user as a sequence of text and use sequence encoder blocks such as transformers to transform that sequence into a fixed-sized representation.

We take a similar approach inspired by \cite{Hajjar2021,Nestler2021} which leverages previous purchases as inputs. 
In our work, we explore how to use other types of user interactions such as Add2Bag to increase user coverage. 
Besides that, we also introduce feedback from returned purchases.

Including other types of customer data such as weight or height to deal with the problem of cold-start recommendation is studied in \cite{Lefakis2020}.
Instead, we focus on increasing the number of information available on the customer's sequence as a way to increase customer coverage.

\section{Methodology}
We approach the problem of size recommendation as a classification task, where given some user and product information $x = (x_{user}, x_{product})$ we want to predict a probability distribution across the feasible sizes $p(y|x)$.
A model $f(x)$ whose output is the distribution $p(y|x)$ can be trained using a classification loss such as Cross-Entropy.

The cardinality of the sizes available in a marketplace can be large, making the sparsity of interactions between users and product variants very high.
A product variant is the combination of the pair (product, size).
Even though we have a large number of sizes, they are typically grouped inside a scale in an ordered manner.
If we look at the position of a size inside the scale (size position), instead of thousands of sizes we can have dozens of positions.
For example, two different sizes "\textit{Gucci EU S}" and "\textit{Gucci IT 46}" share the same relative position (i.e 1) in their respective scales\textit{"Gucci EU"} and "\textit{Gucci IT}".
This means, that instead of predicting a possible size across all the available sizes in the marketplace, we can predict the relative position of the size in the corresponding scale.

To overcome the issue of having recommendations for users that didn't appear at training time, we create a new family of models named Sequence Size Predictor (SSP), these models are inspired by sequence-based recommendation systems \cite{Quadrana2019}.
 Instead of seeing each user ($x_{user}$) as a simple \textit{user\_id}, we use a sequence of events.
You can think of it as $x_{user} = (e_1, e_2, ..., e_T)$, where $e_n$ is a structured textual representation of each event field.
This representation is easily transformed into a categorical multivariate time series by encoding each field as integers: $e_n \in \mathbf{R}^d$, where $d$ is the number of fields of each event.
The events are user interactions within the marketplace, such as \textit{Order, Add2Bag, Add2Wishlist, etc}.


Using such a framework is simple to use negative feedback from the users such as returned purchases.
We do this by adding a field in the \textit{Order} events named \textit{return\_reason}. 
This field indicates the reason why the order might have been returned, for example: \textit{too large}, \textit{not returned}, or \textit{too small}.

By using such a representation of the user, we are no longer depending on seeing the user in training to make a prediction.
Our only requirement is that the user has at least 1 event in the \textit{Event History}.
In the next sections, we will explain how to transform the sequence of user events $x_{user}$ in a fixed-size vector representation that can be used in a classification model.

\subsection{SSP-LSTM}
\begin{figure}[h]
    \centering
    \includegraphics[width=1\textwidth]{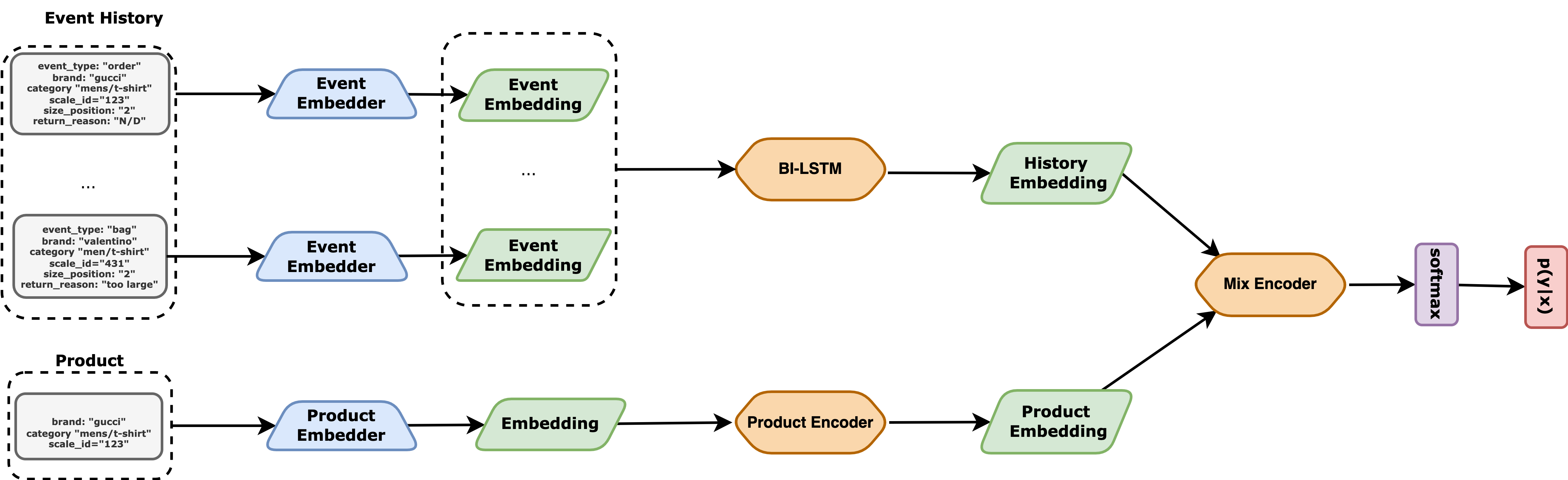}
    \caption{SSP-LSTM architecture}
    \label{fig:ssp-lstm}
\end{figure}

In the first model, we use a Bi-directional LSTM (Bi-LSTM) to encode the sequence of events, the whole architecture can be seen in figure \ref{fig:ssp-lstm}.

We start by using an Embedding Layer to transform each of the event fields into a vector representation called "Event Embedding".
If we assume that each embedding's field is of dimension $v$, each event on the sequence is transformed into a vector of size $\#fields \times v$.
The sequence of event embeddings is passed through a Bi-LSTM and the last hidden state is kept as the "History Embedding", which is a condensed fixed-size representation of the user.
The same Embedding Layer used in the events is also used to transform the common fields present in the product input $x_{product}$. 
The resulting embedding is passed through a multilayer perceptron (MLP) named "Product Encoder" and its output is used as "Product Embedding".
Finally, the "History Embedding" and the "Product Embedding" are concatenated and passed through another MLP "Mix Encoder".
Its output is then used as input of a softmax layer to get a valid probability distribution $p(y|x)$.

Architectures of this nature offer several engineering advantages. It becomes feasible to cache the intermediate representation of the "History Embedding" or the "Product embedding", resulting in potential time savings during prediction.

\subsection{SSP-Attention}

\begin{figure}[h]
    \centering
    \includegraphics[page=1,width=1\textwidth]{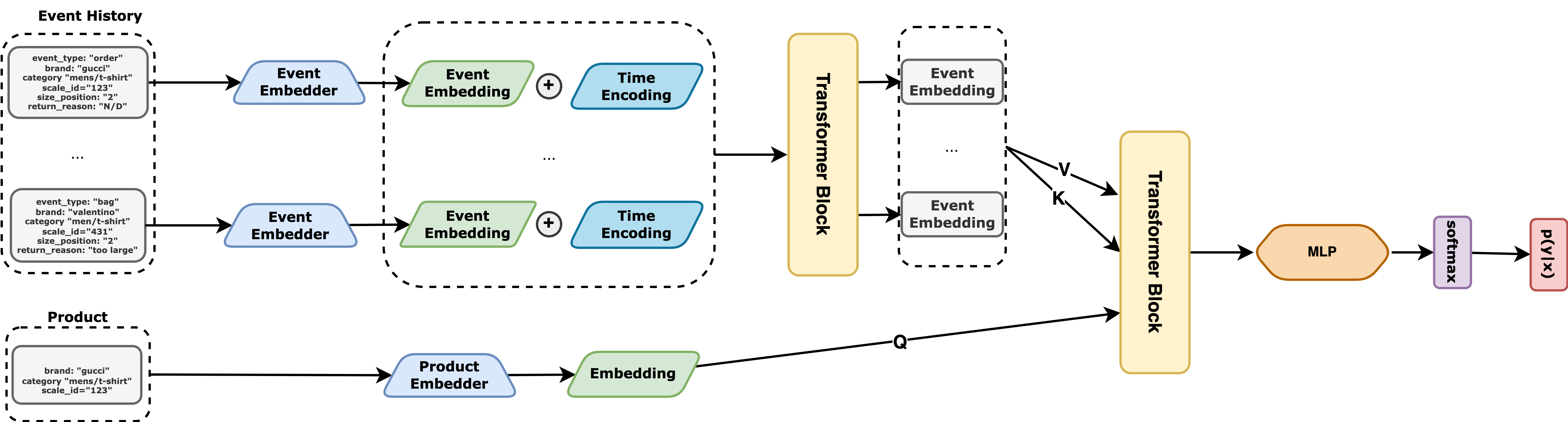}
    \caption{SSP-Attention architecture}
    \label{fig:ssp-attention}
\end{figure}

This model is based on a transformer block \cite{AttentionIsAllYouNeed2017} to encode the event history, it takes inspiration from the attention-based model proposed on \cite{Hajjar2021}.
The whole architecture can be found in figure \ref{fig:ssp-attention}.

Firstly, we embed the sequence of the events in a similar way to what we described before for the SSP-LSTM model.
The only difference is that these embeddings are added a positional encoding based on the number of days that passed since the event occurred.
The sequence of "Event Embedding" is then passed through a first transformer block.
We use the input sequence as query, key, and value which means that the output will be a sequence of embeddings with the same length as the input, we call this sequence the "Transformed Event History".
The same strategy as in the SSP-LSTM is applied to the product input fields which we name "Product Embedding".
Notice that in this model we don't apply an extra MLP to the resulting embedding.

To get a fixed-sized vector, we combine the "Transformed Event History" sequence with the "Product Embedding" by using another transformer block.
In this transformer block the key and the value is the "Transformed Event History" sequence and the query is a sequence only with the "Product Embedding", resulting in an output sequence always with length 1.
Finally, the output of the last transformer block is passed through an MLP and a softmax layer giving the final output distribution.

\section{Experiments}


\subsection{Baselines}

\subsubsection{Personalized Most Common Value (PMCV)}

The PMCV model is a simple rule-based model, it can be seen as a personalized popularity baseline.
This baseline is inspired by the baselines proposed on \cite{Eshel2021}.
The model outputs the most common size for a predefined set of keys seen in the training data. 
These predefined keys are defined in a hierarchical manner, from more specific (i.e. \textit{user\_id, category\_id, brand\_id}) to less specific (i.e \textit{brand\_id}).

At training time, we save a dictionary containing the most common sizes seen across each key.
At inference time, we get the value for the more specific key that is available.

\subsubsection{SFNet}

SFNet is an embedding-based model proposed in \cite{Sheikh2019}.
It learns a representation for each user and another for each product. 
These representations are jointly encoded using an MLP with many hidden layers, and a final output distribution across the different sizes is computed by using softmax.

To compute the user representation ($x_{user}$), we use the \textit{user\_id} which means we have an embedding vector for each user.
To compute the product representation ($x_{product}$), we use the \textit{product\_id}, \textit{brand\_id}, \textit{category\_id}, and the \textit{scale\_id}. 
An embedding is computed for each of the inputs and they are then combined into a single embedding that represents the product by using a MLP.

\subsection{Experimental Setting}

To train the models, we used historical data of Clothing purchases from our marketplace, excluding returns. We use the size ordered by the user as the label.
Notice that even though the training instances don't have return information the event history for the SSP models has events with returned orders, as shown in figure \ref{fig:ssp-lstm} and \ref{fig:ssp-attention}.

Our data ranges from 2022-01-01 to 2023-01-01 and contains $\approx$3.2M purchases from $\approx$400k different products, and $\approx$750k different users.
We split the data into train/val/test (80\%, 10\%, 10\%) according to a back-testing strategy to maintain causality and prevent data leakage.

SFNet was trained using a batch size of 8128 and Adam optimizer \cite{kingma2015adam} ($lr=1\mathrm{e}{-3}$, $wd=1\mathrm{e}{-5}$, $\beta_1=0.9$, $\beta_2=0.999$) for 35 epochs with early stopping based on the validation set top1 accuracy.

SSP models were trained using a batch size of 1024 and Adam optimizer ($lr=1\mathrm{e}{-3}$, $wd=1\mathrm{e}{-4}$, $\beta_1=0.9$, $\beta_2=0.999$) for 25 epochs with early stopping.
All SSP models use an event history with a maximum of 20 events, the events present in the history are \textit{Order} and \textit{Add2Bag}.

\subsection{Model Performance Comparison}

\subsubsection{Model Evaluation}

We evaluate the models in 4 different scenarios:
\begin{itemize}
    \item General: no filtering is applied to the test set
    \item Multiple Gender: test set is filtered to only contain a subset of users that have bought products from multiple genders.
    \item VIP: test set is filtered to only contain power users, a subset of users that have bought more than the average user in the past.
    \item Training Users: test set is filtered to only contain a subset of users that were present in the training set.
\end{itemize}

The results of the different models on these scenarios can be seen in table \ref{table:all_results}.

\begin{table}[h]
\begin{adjustbox}{width=1\columnwidth,center}
\begin{tabular}{l|ccc|ccc|ccc|ccc|}
\multicolumn{1}{c|}{\multirow{2}{*}{\textbf{model}}} & \multicolumn{3}{c|}{\textbf{General}}                                                                  & \multicolumn{3}{c|}{\textbf{Multiple Gender}}                                                          & \multicolumn{3}{c|}{\textbf{VIP}}                                                                      & \multicolumn{3}{c|}{\textbf{Training Users}}                                                           \\
\multicolumn{1}{c|}{}                                & \multicolumn{1}{c|}{\textbf{top1\_acc}} & \multicolumn{1}{c|}{\textbf{top2\_acc}} & \textbf{top3\_acc} & \multicolumn{1}{c|}{\textbf{top1\_acc}} & \multicolumn{1}{c|}{\textbf{top2\_acc}} & \textbf{top3\_acc} & \multicolumn{1}{c|}{\textbf{top1\_acc}} & \multicolumn{1}{c|}{\textbf{top2\_acc}} & \textbf{top3\_acc} & \multicolumn{1}{c|}{\textbf{top1\_acc}} & \multicolumn{1}{c|}{\textbf{top2\_acc}} & \textbf{top3\_acc} \\ \hline
PMCV                                                 & \multicolumn{1}{c|}{0.328}              & \multicolumn{1}{c|}{0.504}              & 0.606              & \multicolumn{1}{c|}{0.305}              & \multicolumn{1}{c|}{0.455}              & 0.535              & \multicolumn{1}{c|}{0.344}              & \multicolumn{1}{c|}{0.501}              & 0.574              & \multicolumn{1}{c|}{0.353}              & \multicolumn{1}{c|}{0.473}              & 0.517              \\
SFNet                                                & \multicolumn{1}{c|}{0.387}              & \multicolumn{1}{c|}{0.660}              & 0.821              & \multicolumn{1}{c|}{0.378}              & \multicolumn{1}{c|}{0.651}              & 0.815              & \multicolumn{1}{c|}{0.425}              & \multicolumn{1}{c|}{0.707}              & 0.855              & \multicolumn{1}{c|}{0.449}              & \multicolumn{1}{c|}{0.727}              & 0.872              \\
SSP-LSTM                                             & \multicolumn{1}{c|}{0.538}              & \multicolumn{1}{c|}{0.803}              & 0.916              & \multicolumn{1}{c|}{0.479}              & \multicolumn{1}{c|}{0.749}              & 0.881              & \multicolumn{1}{c|}{0.489}              & \multicolumn{1}{c|}{0.764}              & 0.893              & \multicolumn{1}{c|}{0.523}              & \multicolumn{1}{c|}{0.794}              & 0.911              \\
SSP-Attention                                        & \multicolumn{1}{c|}{\textbf{0.564}}     & \multicolumn{1}{c|}{\textbf{0.815}}     & \textbf{0.922}     & \multicolumn{1}{c|}{\textbf{0.505}}     & \multicolumn{1}{c|}{\textbf{0.762}}     & \textbf{0.890}     & \multicolumn{1}{c|}{\textbf{0.508}}     & \multicolumn{1}{c|}{\textbf{0.776}}     & \textbf{0.901}     & \multicolumn{1}{c|}{\textbf{0.548}}     & \multicolumn{1}{c|}{\textbf{0.808}}     & \textbf{0.918}    
\end{tabular}
\end{adjustbox}
\caption{Model Accuracy in the different scenarios.}
\label{table:all_results}
\end{table}

All the models show a better performance than the PMCV baseline, which shows the importance of ML models vs heuristics in the size recommendation problem in a luxury marketplace.
The SSP-Attention model performs better than all the other models across all the different scenarios.

The SSP models show better results than SFNet in all the scenarios.
However, it's interesting to see that on VIP users the SSP models don't show the same trend as SFNet, behaving worse than in the general scenario.
This might be due to the fact that since VIP are power users, they interact more with the marketplace and that might make the need to use a bigger event history.

In all the models, we see a decrease in performance on the MultipleGender vs General scenario.
This is due to the fact that these users are more challenging since they have purchased for different genders under the same account, making it more difficult to predict the underlying size of the customer.

Finally, in the Training Users scenario, we see that SSP models still beat SFnet and PMCV.
It is also important to note that by filtering the users we are reducing the coverage of users to 49.27\% of the users present in the General scenario.
Remember that in this scenario we are guaranteed to have an embedding that was specifically tuned to the user at training time, this explains the better performance of SFnet vs the General scenario.



\subsubsection{Influence of \textit{Add2Bag} events}

In this section, we see the effect of using \textit{Add2Bag} events in the SSP models.
To do this ablation we use a subset of the data that only has users that had both \textit{Add2Bag} and \textit{Order} events in the range of the test set.

We split the usage of the \textit{Add2Bag} events in two different ways: (1) present in the event history at inference time; (2) present in the event history at training time.
These two ways can even be mixed, i.e, we can have a model that was trained with \textit{Add2Bag} events in the event history and at inference time also has \textit{Add2Bag} events in the event history of the instances for which is predicting.
In table \ref{table:ablation_bag_ssp}, you can see the results for the SSP models for all 4 different combinations:

\begin{enumerate}
    \item \textit{Add2Bag} in event history \& trained only with \textit{Order} in event history.
    \item \textit{Add2Bag} in event history \& trained with \textit{Order} and \textit{Add2Bag} in event history.
    \item no \textit{Add2Bag} in event history \&  trained only with \textit{Order} in event history.
    \item no \textit{Add2Bag} in event history \& trained with \textit{Order} and \textit{Add2Bag} in event history.
\end{enumerate}

\begin{table}[h]
    \begin{tabular}{lc|ccc|ccc}
    \multicolumn{1}{c}{\multirow{2}{*}{\textbf{model}}} & \multirow{2}{*}{\textbf{\begin{tabular}[c]{@{}c@{}}add2bag in\\ event history\end{tabular}}} & \multicolumn{3}{c|}{\textbf{\begin{tabular}[c]{@{}c@{}}trained only with order\\  events in history\end{tabular}}}        & \multicolumn{3}{c}{\textbf{\begin{tabular}[c]{@{}c@{}}trained with add2bag \& orders\\ events in history\end{tabular}}}  \\
    \multicolumn{1}{c}{}                                &                                                                                              & \multicolumn{1}{l}{\textbf{top1\_acc}} & \multicolumn{1}{l}{\textbf{top2\_acc}} & \multicolumn{1}{l|}{\textbf{top3\_acc}} & \multicolumn{1}{l}{\textbf{top1\_acc}} & \multicolumn{1}{l}{\textbf{top2\_acc}} & \multicolumn{1}{l}{\textbf{top3\_acc}} \\ \hline
    SSP-LSTM                                            & yes                                                                                          & 0.451                                  & 0.738                                  & 0.887                                   & 0.528                                  & 0.797                                  & 0.913                                  \\
    SSP-Attention                                       & yes                                                                                          & 0.550                                  & 0.809                                  & 0.919                                   & \textbf{0.554}                         & \textbf{0.810}                         & \textbf{0.919}                         \\
    SSP-LSTM                                            & no                                                                                           & 0.496                                  & 0.771                                  & 0.901                                   & 0.498                                  & 0.772                                  & 0.901                                  \\
   SSP-Attention                                       & no                                                                                           & 0.515                                  & 0.780                                  & 0.904                                   & 0.513                                  & 0.779                                  & 0.904                                 
    \end{tabular}
\caption{Effect of \textit{Add2Bag} events in SSP models.}
\label{table:ablation_bag_ssp}
\end{table}

For both models, we see that the best results are when we use \textit{Add2Bag} and \textit{Order} events in the event history at training time and at inference time.

We can also see that for the SSP-LSTM model if we don't have \textit{Add2Bag} events at training time and only use them at inference we have worse results.
In the SSP-Attention model, we see that we can have good accuracy by only adding \textit{Add2Bag} events at inference time. 

Besides this accuracy improvement, by using \textit{Add2Bag} events we see an increase of 24.5\% in the users that can receive a size recommendation.
Remember that to make a recommendation our requirement is that the user has at least 1 event in history.

\subsubsection{Influence of \textit{return\_reason} field}

We present in table \ref{table:return_reason_ablation} an ablation regarding the effect of using or not using the \textit{return\_reason} field in the SSP models.
This ablation was done using a dataset only with \textit{Order} events in the event history.

\begin{table}[h]
\begin{tabular}{lc|c|c|c}
\multicolumn{1}{c}{\textbf{model}} & \textbf{return\_reason} & \textbf{top1\_acc} & \textbf{top2\_acc} & \textbf{top3\_acc} \\ \hline
SSP-LSTM                           & yes                     & 0.497              & 0.771              & 0.901              \\
SSP-Attention                      & yes                     & \textbf{0.515}              & \textbf{0.781}              & \textbf{0.905}              \\
SSP-LSTM                           & no                      & 0.494              & 0.767              & 0.898              \\
SSP-Attention                      & no                      & 0.503              & 0.769              & 0.898             
\end{tabular}
\caption{Effect of using \textit{return\_reason}.}
\label{table:return_reason_ablation}
\end{table}

We see that for the SSP-Attention model, we have an increase of 2.33\% in accuracy, while in the SSP-LSTM model the increase is negligible.


\subsection{Model Latency Comparison}

In a real production use case, an essential performance metric to consider is model latency, which measures the time the model takes to make a prediction (or a batch of predictions) in milliseconds.

\begin{figure}[h]
    \centering    
    \subfloat[\centering Average model latency Vs batch size]{\includegraphics[width=0.45\textwidth]{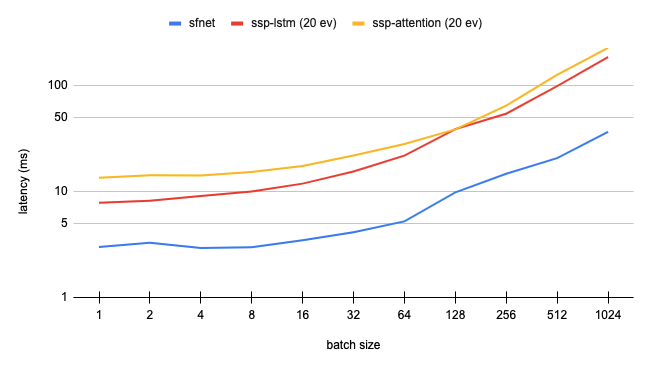}}%
    \qquad
    \subfloat[\centering Average model latency Vs event history size]{\includegraphics[width=0.45\textwidth]{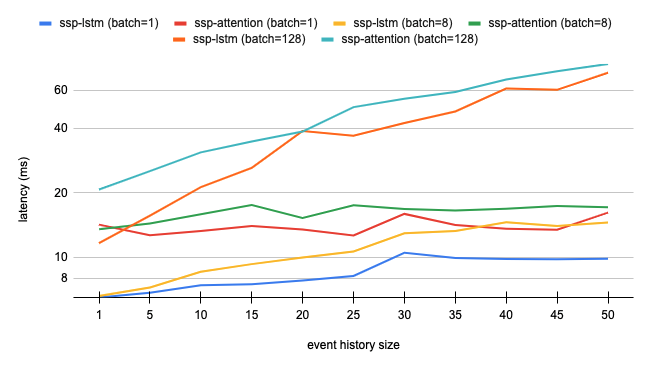}}%
    \caption{Model Latency Benchmark}%
    \label{fig:model_latency_benchmark}%
\end{figure}

Figure \ref{fig:model_latency_benchmark} illustrates how the average model latency changes with the batch size and the length of the event history using synthetic data and models with consistent hyperparameters as described in the Experimental Setting section. The inference is performed on a CPU (\textit{11th Gen Intel® Core™ i7-11800H @ 2.30GHz × 16}), which better reflects production machine types.

Figure \ref{fig:model_latency_benchmark} (a) shows that SFNet consistently exhibits lower latency than the SSP models across all batch sizes, primarily because SFNet processes smaller inputs (without sequences of events) and has fewer computations. For the SSP models, only results with 20 events in the history are presented.

The comparison between SSP-LSTM and SSP-Attention reveals that SSP-LSTM consistently has lower latency due to the extra computation involved in the 2 attention layers of SSP-Attention. Both models achieve a latency of under 50ms for batch sizes smaller than 128. In production, a batch size of 1 is commonly used, and all models have a latency well below 20ms, respectively 2.99 ms for SFNet, 7.81 ms for SSP-LSTM, and 13.51 ms for SSP-Attention.

Figure \ref{fig:model_latency_benchmark} (b) examines how model latency changes with the size of the event history for various batch sizes. For larger batch sizes (128), latency increases with an increase in event history size. Interestingly, latency appears to stabilize for smaller batch sizes after 30 events in history. The insight here is that this behavior is possibly linked to torch optimization, mainly when batching a limited number of events when making fewer predictions (batch size = 1 or 8).

\section{Conclusion}
In this work, we showed how size recommendations can be solved in a luxury fashion marketplace using only historical purchase data.

By using SSP models with an attention mechanism (SSP-Attention), we achieve an improvement of 72.2\% in top1 accuracy compared to a rule-based heuristic, 45.7\% compared to SFNet \cite{Sheikh2019} and 4.8\% compared to SSP-LSTM.
We also show the importance of using \textit{Add2Bag} interactions to increase both the accuracy and coverage of the customers.
Finally, we show that SSP models besides increasing the accuracy are still capable of having a latency below ~15ms which is still usable in production.

For future work, we would like to explore ways of improving the SSP models by including more information regarding the product, such as product measurements. 
Including more information regarding the customer such as measurements, body shape, and other implicit signals besides Add2Bag is also an open topic.


\bibliographystyle{ACM-Reference-Format}
\bibliography{main}


\end{document}